\documentclass[epj]{svjour}
\usepackage{graphics}

\usepackage{subcaption}
\usepackage[utf8]{inputenc} 
\usepackage[T1]{fontenc}    
\usepackage{hyperref}       
\usepackage{url}            
\usepackage{booktabs}       
\usepackage{amsfonts}       
\usepackage{nicefrac}       
\usepackage{microtype}      
\usepackage{lipsum}		
\usepackage{graphicx}
\usepackage{natbib}
\usepackage{doi}
\usepackage{amssymb}
\usepackage{xcolor}

\usepackage{amsmath}
\usepackage{comment}
\begin{document}
\title{Angular Correlation Function from sample covariance with BOSS and eBOSS LRG}
\author{Paula S. Ferreira\inst{1}
\thanks{\emph{Present address:} psfer@pos.if.ufrj.br}
\and Ribamar  R. R. Reis\inst{1,2}
\thanks{\emph{Present address:} ribamar@if.ufrj.br}%
}                     
\offprints{}          
\institute{Instituto de Física, Universidade Federal do Rio de Janeiro
	Rio de Janeiro, RJ, Brazil CEP 21941-972  \and Observatório do Valongo, Universidade Federal do Rio de Janeiro, Rio de Janeiro, RJ, Brazil CEP 20080-090}
\date{Received: date / Revised version: date}
%
\abstract{
The Baryon Acoustic Oscillations (BAO) are one of the most used probes to understand the accelerated expansion of the Universe. Traditional methods rely on fiducial model information within their statistical analysis, which may be a problem when constraining different families of models. 
This work aims to provide a method that constrains $\theta_{BAO}$ through a model-independent approach using the covariance matrix from the galaxy sample from thin redshift bins, later validated with a mock sample covariance matrix. We used widths of $\delta z = 0.002$ separation for all samples as the basis for a sample covariance matrix weighted by the statistical importance of the redshift bin. Each sample belongs to the Sloan Digital Sky Survey: BOSS1, BOSS2, and eBOSS, with effective redshift $z_{eff}$: 0.35, 0.51, 0.71, and different numbers of bins with 50, 100, and 200. 
To get $\theta_{BAO}$, we correct the angular separation from the polynomial fit ($\theta_{fit}$) by comparing each bin correlation function with the correlation function of the whole set, a parameter named $\tilde{\alpha}$. We also tested such correction by choosing the bin at $z_{eff}$ and found that for eBOSS $\theta_{BAO}$ is in $1 \sigma$ agreement with the Planck 18 model.
Finally, we found that the sample covariances are noisy compared to the mocks for lower $z$ samples, something expected due to nonlinear effects. Such noise impact can be seen in the parameter constraints but does not affect the eBOSS covariance sample. It is shown that mocks' results do tend to its chosen fiducial cosmology $\theta_{BAO}$. BOSS1 and BOSS2 showed agreement with Planck 18 and an agreement with Pantheon + S$H_0$ES when $\tilde{\alpha}$ is based on the bin $z=z_{eff}$.
\PACS{
      {98.80.Es}{Observational Cosmology}  
     } 
} 

\authorrunning{When $\tilde{\alpha}$ is based on the bin $z=z_{eff}$ we found bins in agreement with Planck 18.

Paula Ferreira\and 
Ribamar Reis}
\titlerunning{2pacf from sample covariance}

\maketitle
\section{Introduction}
The Baryon Acoustic Oscillations (BAO) are one of the most used probes to understand the accelerated expansion of the Universe. Cosmological information can be extracted through the two-point correlation function and power spectra estimated with the sky distribution and redshift of standard tracers \cite{peebles2001galaxy}. Among the tracers, the most used are the luminous red galaxy (LRG)  first used by \cite{eisenstein2005detection}  and \cite{percival20012df}. Now, a multi-tracer analysis is possible with emission line galaxies \citep{wang2020clustering,de2021completed}, quasars \cite{hou2021completed} and Lyman-$\alpha$ forests \citep{des2020completed}. 

Future and current surveys will reach a larger number of observed objects, such as the Dark Energy Spectroscopic Instrument (DESI) \cite{flaugher2014dark},  the Dark Energy Survey (DES) \cite{rosell2022dark}, the Large Synoptic Survey Telescope (LSST) survey\cite{ansari2019impact}, the Javalambre-Physics of the Accelerated Universe Astrophysical Survey (J-PAS) \citep{benitez2014j}, Euclid \citep{scaramella2022euclid}. Larger samples have the advantage of being statistically robust and less susceptible to cosmic variance \citep{moster2011cosmic}.  Their goal is to reach higher precision in order to test different cosmological models. This is only achievable through a template analysis that can be applied to any model. 

The traditional methods rely on fiducial cosmology and \textit{ad hoc} parameters to fix nonlinear effects \citep{seo2008nonlinear}. Moreover, the statistical analysis carries fiducial model information within its template. A possible issue with these methods is the applicability to test other families of models. \cite{sanchez2011tracing, sanchez2013precise} analysed the angular and radial correlation functions using a polynomial fit, also independent of the model. \cite{marra2019first} and \cite{menote2022baryon} followed this construction using the Sloan Digital Sky Server (SDSS) precise surveys. Other alternative approaches have been proposed in the literature such as the new standard ruler called Linear Point \citep{anselmi2016beating,anselmi2018linear}, which can be used to derive a purely geometrical BAO distance \citep{anselmi2019cosmic}.

Parameter estimation from Large Scale Structure (LSS) galaxy counting requires a covariance matrix. Usually, the community uses hundreds of mock catalogs that mimic surveys. This can be done by N-body simulations such as the MultiDark-Patchy Mocks by \citep{kitaura2016clustering,rodriguez2016clustering}, and the N-body Parallel Particle-Mesh GLAM code (PPMGLAM) by \cite{klypin2018dark}, which is the core of GLAM (GaLAxy Mocks). Simulation mocks are highly computationally expensive because they solve the matter density field evolution. The solution for many collaborations was to use a modest mock construction like the Log-Normal mocks, that can be found in \texttt{CoLoRe} \citep{ramirez2022colore}, \texttt{FLASK} \citep{xavier2016improving}, \texttt{nbodykit} \citep{hand2018nbodykit} \texttt{LogNormalCatalog}. Both approaches require fiducial cosmology, something that would be desirable to avoid.

\cite{scranton2002analysis} estimated the covariance matrix assuming that the error distribution is Gaussian. \cite{zehavi2002galaxy,zehavi2004departures} used the so-called jackknife error estimate, by dividing a larger sample into sub-samples in order to find the covariance matrix. Their analysis concluded that the results were representative error estimators. \cite{ross20072df} also used sub-samples to find the covariance matrix and error bars from the data set. However, there was a lack of observed objects. The latest data sets from spectroscopic surveys must provide even more representative errors and covariance matrices once we are provided with more objects, thus thinner redshift sub-samples.

This work aims to provide a pipeline to constrain $\theta_{BAO}$ through a model-independent estimation of the angular correlation function using the covariance matrix from the data sample (sample covariance). We use thin bins of redshift with width $\delta z = 0.002$ as the basis for a such matrix weighted by the statistical importance of the redshift bin. Our analysis compares the sample covariance to the covariance matrix originating from mock catalogs and also the traditional covariance from mocks, without the binning. We compare samples with different effective redshift values, from lowest $z$ to highest $z$ and check the consistency according to each scale. Furthermore, we correct the angular separation that encodes the BAO feature of the \cite{sanchez2011tracing} polynomial fit by a bias function that compares the correlation function of each bin with the whole set. 

Our analysis is first described with our methodology in \ref{sec:met} where we describe the data used in \ref{sec:data},  how it was divided into redshift bins in \ref{sec:bins}, and finally its correlation function estimator in \ref{sec:corre}. Next, we discuss the construction of the covariance matrix with the data and with mocks in \ref{sec:cov}. The polynomial function is described in \ref{sec:poly} and our $\theta_{BAO}$ method in section \ref{sec:thetabao}. The results and conclusion are found in \ref{sec:results} and \ref{sec:conclusions}, respectively.
\section{Methods}\label{sec:met}

\subsection{Data and mocks}\label{sec:data}

In this work, we used two data sets from the SDSS. The LRG and the LRG CMASS data from the Data Release 16 (DR16) \cite{wang2020clustering} of the extended Baryon Oscillation Spectroscopic Survey (eBOSS) spectroscopic observations from the SDSS IV. The sample is distributed in Northern Galactic Cap (NGC) and Southern Galactic Cap (SGC) with a redshift range of $0.6\leq z \leq 1.0$ and a total of $552,274$ galaxies. Its random catalogue contains fifty times more galaxies than the real one.

The other set consists of galaxies from the BOSS sample of the SDSS-III DR12. We separated a redshift range $0.3\leq z<0.4$ called the BOSS1 set and BOSS2 range is $0.4\leq z<0.6$. The summary of the data specification can be found in Table~\ref{tab:data}. We chose this minimum redshift because, according to our tests, smaller redshift ranges did not show a significant signal for the BAO feature, as expected.

For the eBOSS set, we used $N=1,000$ realisations of realistic mocks catalogs from  \cite{zhao2021completed}, for each galactic cap, based on the effective Zel'dovich approximacovariancetion mock generator (EZmock) by \cite{chuang2015ezmocks}. These mocks were constructed using the Zel'dovich approximation \cite{zel1970gravitational} which is faster than producing mocks using N-body simulations and accurate in clustering statistics. 

The BOSS set was validated with $N=1,000$ realisations of the MultiDark-Patchy mocks (MD-Patchy mocks). These mocks were constructed for both the LOWZ redshift range \cite{kitaura2016clustering} and the CMASS part \cite{rodriguez2016clustering}. We must keep in mind there is a difference between the two mocks' construction, MD-Patchy mocks were written from the N-body simulation BigMultiDark simulations \cite{klypin2016multidark} using the Patchy code \cite{kitaura2014modelling,kitaura2015constraining} which considers a biasing model. EZmock also used BigMultiDark, but their approach is not fully N-body based, and different from PATCHY, they avoid a strict biasing model. 

We cut the mocks to match the survey's redshift range. The same routine was applied to calculate the angular correlation function for the mocks using the same random catalogue for all mocks.

\subsection{Redshift bins}\label{sec:bins}

In order to get a direct BAO measurement, we used an approach similar to the one developed by \cite{sanchez2011tracing}. 

eBOSS sample was divided into 200 redshift bins with width $\delta z = 0.002$. The choice was made considering the distribution of galaxies in the sky. One wants many bins to construct a fair covariance matrix and enough galaxies in each bin to get the BAO feature through angular counting. The angular pairs of each bin were computed using the code \texttt{corrfunc} developed by \cite{corrfunc}. The data-data ($DD$), the data-random ($DR$), and random-random ($RR$) pairs counts calculated for angular pixels from $0.8^{\circ}$ to $10^{\circ}$ with width $\delta \theta \sim 0.4^{\circ}$. This was also a convenient choice to find the BAO feature. Larger angles would provide a smeared function, while smaller ones would have too few galaxies.

\begin{figure}
    \centering    
    \resizebox{0.5\textwidth}{!}{\includegraphics{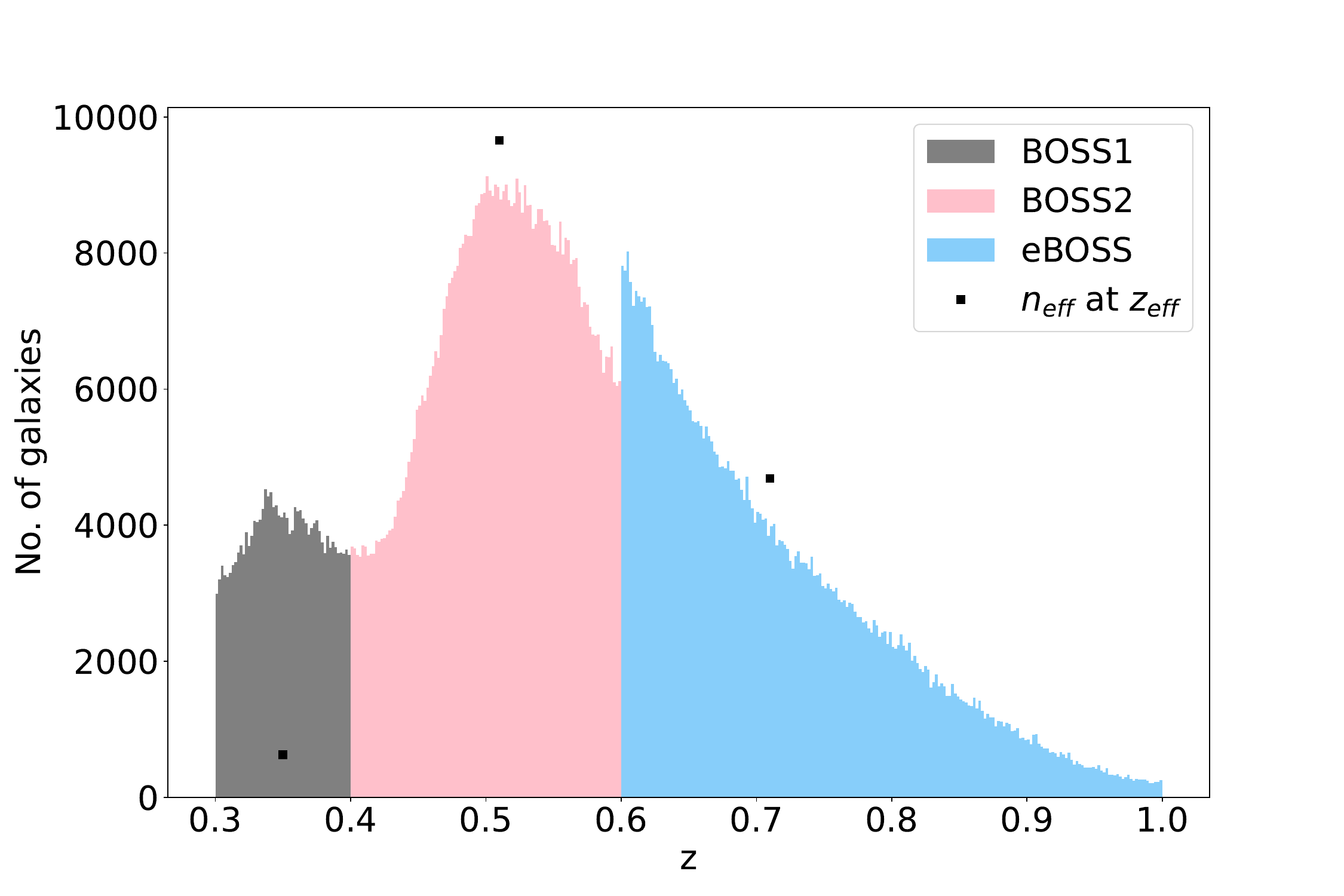}}
    \caption{Each sample bin separation. The black squares are the $n_{eff}$ of each sample. BOSS1, BOSS2, and eBOSS are in gray, pink, and blue, respectively.}
    \label{fig:histogram}
\end{figure}

The BOSS1 set was divided into 50 bins with width $\delta z = 0.002$, angular range $1^{\circ}<\theta<12^{\circ}$ with $\delta \theta \sim 0.31^{\circ}$. BOSS2 was divided into 100 bins with the same redshift separation and the same angular configuration as eBOSS. The redshift distribution for the chosen binning can be visualized in Figure~\ref{fig:histogram}.

With the pairs counts, we computed the angular two-point correlation function for each redshift bin using the Landy-Szalay estimator \cite{landy1993bias}. Each angular function was normalised according to its corresponding i-th bin length. The angular correlation function ($\mathrm{w}_i$) of the i-th bin is written in the following equation:
\begin{equation}
\mathrm{w}_i(\theta)= \left( \frac{N_{i,rand}}{N_{i,data}}\right)^2 \frac{DD_i(\theta)}{RR_i(\theta)} - 2 \frac{N_{i,rand}}{N_{i,data}}\frac{DR_i(\theta)}{RR_i(\theta)}+1 \text{ ,}
\label{eq:tpcf}
\end{equation}
where $N_{i,rand}$ is the number of galaxies in the random bin, and $N_{i,data}$ in a data bin. 
\begin{table}
\centering
\scriptsize	
	\caption{Data specifications}
	\label{tab:data}
	\begin{tabular}{lllll} 
		\hline\noalign{\smallskip}
            
		Sample & $z$ range & \# of Galaxies & $z_{eff}$ & No. of bins\\
            \noalign{\smallskip}\hline\noalign{\smallskip}
            
		BOSS1 & $0.3\leq z<0.4$ & $192,285$ & $0.35$ & $50$\\
		BOSS2 & $0.4\leq z<0.6$ & $686,370$ & $0.51$ & $100$\\
		eBOSS & $0.6\leq z \leq 1.0$ & $552,274$ & $0.71$ & $200$\\
		\noalign{\smallskip}\hline
	\end{tabular}
\end{table}

\subsection{Weighting-scheme and correlation function}\label{sec:corre}

In addition to normalisation, it is important to account for the statistical significance of each bin. Our approach was to use the total weights ($w_{tot,a}$) that account for the systematic effects of the spectrographs. For the LRG sample, the weights given by
\begin{equation}\label{eq:weight0}
    w^{LRG}_{tot}=w_{sys,a}\cdot w_{cp,a}\cdot w_{noz,a},
\end{equation}
which is described in \cite{gil2020completed}. We used the prescription described in \cite{reid2016sdss} with the pCMASS, eBOSS and CMASS BOSS samples:
\begin{equation}\label{eq:weight}
    \displaystyle w_{tot,a}=w_{sys,a}(w_{cp,a}+w_{noz,a}-1),
\end{equation}
where the total galaxy weights $w_{tot,a}$ depends on the total angular systematic weights $w_{systot,a}$, the weight for close pair correction $w_{cp,a}$, and the total weight to the nearest neighbour of the redshift failure $w_{noz,a}$. \texttt{corrfunc} \citep{corrfunc} weights the pairs using the \texttt{pair\_product} method, which is simply multiplying the weights of the galaxies in a pair.

We computed the effective redshift of the three samples according to the following equation based on \cite{beutler20116df}:
\begin{equation}
    z_{eff} = \frac{\displaystyle \displaystyle \sum_{a<b} \frac{w_a w_b(z_a+z_b)}{2}}{\displaystyle \displaystyle \sum_{a<b}w_a w_b} \text{ ,}
\end{equation}
where $a$ and $b$ represent galaxies in a sample, and $w_{a | b}$ is their $w_{tot}$. The values are shown in Table~\ref{tab:data}.

Random and mock bins do not carry the systematic effects of fibres on the spectrographs. Since the available FKP weight by \cite{feldman1993power} accounts for the number of galaxies in a volume, we use unit weights $w=1.0$ for the angular correlation function. Although EZmock contains synthetic systematics within their weights we kept them unitary as a matter of consistency with MD-Patchy mocks.

As the bins have different sizes and redshifts, the bins also need to be weighted. For that, we consider a bin weighting revisiting \cite{marra2019first}. Now, redshift bins are weighted instead of pixels.

The weighting of each bin is similar to the variance of the weights of each random galaxy. For each bin, the weight $rr_{i}$ is written as:
\begin{equation}\label{eq:weight_rr}
    rr_{i}=\frac{\left(\displaystyle \displaystyle \sum_{j} RR_{ij}\right)^2-\displaystyle \displaystyle \sum_j RR_{ij}^2 }{2}
\end{equation}
where $i$ stands for the redshift bin and $j$ for $j$-angular separation. \cite{marra2019first} wrote $rr_\alpha$ as a weight for each pixel; here we use the pair counting $RR$ because all of them are uncorrelated with the observed points but still an accurate representation of them in terms of redshift. We want to weigh the bins according to statistical significance and that is related to how many galaxies are in the $ij$ box. 

Finally, the correlation function $\mathrm{W}(\theta)$ is computed as a weighted mean of the functions for each bin \begin{equation}\label{eq:CF}
    \mathrm{W}(\theta)=\frac{\displaystyle \displaystyle \sum\limits_i \mathrm{w}_i(\theta) \cdot rr_{i}}{\displaystyle \displaystyle \sum\limits_i rr_{i}}.
\end{equation}

\section{Covariance matrix and \texorpdfstring{$\theta_{BAO}$}{Lg} }\label{sec:cov}

The whole sample has an effective number of galaxies per bin ($n_{eff}$):
\begin{equation}\label{eq:neff}
    n_{eff}  = \frac{\left(\displaystyle \sum_i rr_i\right)^2}{\displaystyle \sum_i rr_i^2}.
\end{equation}
We consider a good sample the one that has an $n_{eff}$ close to the number of galaxies at the $z_i \sim z_{eff}$. This can be seen in Figure~\ref{fig:histogram}, where we display the $n_{eff}$ of each sample as black squares. The BOSS2 and eBOSS present this characteristic, while the smaller sample with fewer bins shows disagreement with the size of the bin. 

Equation~(\ref{eq:tpcf}) works as a matrix of many correlation function realisations, similar to the usual usage of mocks.
We use each $\mathrm{w}_i(\theta)$ compared to $\mathrm{W}(\theta)$ from equation \ref{eq:CF} to construct the sample covariance matrix $ \zeta_{ml}$:
\begin{align}\label{eq:free_cov}
    \zeta_{ml}=\frac{n_{eff}^{-1}}{(N_i-1)} \displaystyle \sum\limits_{i=1}^{N_i} [ \mathrm{w}_i(\theta_m) - \mathrm{W}(\theta_m)] \nonumber \\ \times [\mathrm{w}_{i}(\theta_l)-\mathrm{W}(\theta_l)]. 
\end{align}
Basically, we compute the bias between each $\mathrm{w}_i(\theta)$ and $\mathrm{W}(\theta)$, but because $\mathrm{w}_i(\theta)$ is not weighted according to Eq.~(\ref{eq:weight}), so we must correct this by $n_{eff}$.

In order to validate our method we require realisations of the survey catalogues from mock catalogues. Mocks are an approximation of a survey concerning its footprint and redshift. However, when analysing the clustering statistics we need to remember that mocks have the challenge of accounting for nonlinear effects which are inevitable for lower redshift values, which is mentioned by other studies.
We use mocks to compute $\mathrm{W}_k(\theta)$ of the $k$-mock and test the method. Now, the covariance matrix $C_{ml}$:
\begin{align}
    C_{ml} = \frac{n_{eff}^{-1}}{N-1} \displaystyle \sum\limits_{k=1}^{N=1000} [ \mathrm{W}_k(\theta_m) - \mathrm{\bar{W}}(\theta_m)] \nonumber \\ \times [\mathrm{W}_{k}(\theta_l)-\mathrm{\bar{W}}(\theta_l)],\label{eq:covmocks}
\end{align}
where $\mathrm{\bar{W}}$ is the average over the correlation function $W_k(\theta)$ of each mock.
Nevertheless, Eq.~(\ref{eq:covmocks}) is not model-independent, because the mocks were constructed assuming a fiducial cosmology.

\subsection{Polynomial function}

We used \cite{sanchez2011tracing} expression below for the angular two-point correlation function to fit our estimated $W(\theta)$.
\begin{equation}
    \mathcal{W}(\theta) = A+B\theta^\gamma+C e^{-\frac{(\theta-\theta_{fit})^2}{2\sigma^2}}
\end{equation}
The parameter $A$ is related to the behaviour of the function after the BAO peak. $B$ and $C$ weigh the importance of their respective terms, if there is no BAO peak, for instance, $C=0$. $\gamma$  is the power law of the function's overall shape. The physical parameters are $\theta_{fit}$ and $\sigma$, $\theta_{fit}$ gives the position of the BAO, while $\sigma$ is the width of the BAO.

We applied Gaussian priors to the parameters that are related to the BAO: $C$, $\theta_{fit}$, and $\sigma$, the priors are written in table~\ref{tab:priors}. The parameter estimation was done using the maximum likelihood estimator method through the \texttt{emcee}'s Affine Invariant Markov Chain Monte Carlo (MCMC) Ensemble sampler software by \cite{2013PASP..125..306F}. The fitting is done with the $W(\theta)$ from the data, we will compare the best-fit results alternating between $\zeta_{ml}$ and $C_{ml}$.

\begin{table}
\centering
    \scriptsize	
    \begin{tabular}{llll}

        \hline\noalign{\smallskip}

          & $C$ & $\theta_{fit}$& $\sigma$ \\
          \noalign{\smallskip}\hline\noalign{\smallskip}
            
          BOSS1 &$\mathcal{N}(5\times 10^{-3},1\times 10^{-5})$& $\mathcal{N}(6.00,0.01)$& $\mathcal{N}(1.00,0.01)$\\

          BOSS2 & $C>0$&$\mathcal{N}(4.2,0.1)$ & $\mathcal{N}(0.04,0.01)$\\

          eBOSS & $\mathcal{N}(7.60 \times 10^{-3}, 1\times 10^{-5})$& $\mathcal{N}(2.70,0.01)$ & $\sigma > 0$\\

         \noalign{\smallskip}\hline
    \end{tabular}
    \caption{Priors used in the MCMC, both with $\zeta_{ml}$ and $C_{ml}$.}
    \label{tab:priors}
\end{table}

The chi-squared function is:
\begin{equation}
    \chi^2 = \displaystyle \sum\limits_{ml}\left[ \mathrm{W}(\theta_m)- \mathcal{W}(\theta_m)\right]\zeta_{ml}^{-1}\left[\mathrm{W}(\theta_l)- \mathcal{W}(\theta_l)\right],
\end{equation}
the same equation is used with $C_{ml}$.

\subsection{The BAO scale \texorpdfstring{$\theta_{BAO}$}{Lg} }\label{sec:thetabao}
Should the estimator and its covariance matrix be representative of the angular pairs counting of the LSS, we can constrain cosmological parameters from fitting a model to the data. For that we need to find the BAO signal, $\theta_{BAO}$ as a function of $\theta_{fit}$. Here, we will focus on the use of a totally model-independent procedure, leaving a model that gets the advantage of our estimator for future work.

We propose a similar method to \cite{carvalho2016baryon} to find $\theta_{BAO}$. We take the correlation function estimation for the whole set, as in Eq.~(\ref{eq:tpcf}) without weighting according to the random bins,  to find $\mathrm{w}_{\delta z = 0}(\theta)$ and compute a bias function, the difference between $\mathrm{w}_{\delta z = 0}(\theta)$ and the $\mathrm{w}_i(\theta)$ not normalised by $rr_i$:
\begin{equation}\label{eq:normed_bias}
    \Delta \mathrm{w}_i(\theta)=\mathrm{w}_i(\theta)-\mathrm{w}_{\delta z = 0}(\theta),
\end{equation}
Next, we take this bias function and compute the 20$^{th}$ percentile and the median over the bins to get our shift parameter $\tilde{\alpha}$ as a function of $\theta$:
\begin{equation}\label{eq:alpha}
    \tilde{\alpha}_{p}(\theta) = \frac{\displaystyle \sum_{i}^{N_{p}} \Delta \mathrm{w}_i(\theta)}{N_{p}} \geq \frac{p}{100},
\end{equation}
$\tilde{\alpha}_p$ is normalised by ${\sqrt{\int |\mathrm{w}_i(\theta)-\mathrm{w}_{\delta z = 0}(\theta)|^2 \mathrm{d} \theta}}$, which is the contribution of all dispersion from $\mathrm{w}_{\delta z = 0}$. Here, $p$ stands for percentile, $\tilde{\alpha}_{20}(\theta)$ represents the bias function of a bin that is greater or equal to 20\% of the bias function of the other bins, $\tilde{\alpha}_{50}(\theta)$ is 50\% greater or equal to the other bias functions. In the case of the bias function for the bins was a normal distribution, the average value would be the percentile 50. Knowing a perfect normal distribution is not achieved we also test the percentile 20, which indicates whether the most statistically significant bias is dislocated from the average.

Finally, $\theta_{BAO}$ is the correction of the model-independent $\theta_{fit}$ according to the bias of each with respect to the correlation function if the binning was not applied:
\begin{equation}\label{eq:theta_bao}
    \theta_{BAO}(z_{eff})= \theta_{fit}+ \tilde{\alpha}(\theta_{fit})\theta_{fit}.
\end{equation}
$\tilde{\alpha}(\theta_{fit})$ is the value of the function $\tilde{\alpha}$ at the angle found from the MCMC fitting results, both from sample covariance and mock covariance. We used $\tilde{\alpha}_{20}$ and $\tilde{\alpha}_{50}$ in Eq. (\ref{eq:theta_bao}) and will present the results for both.

We also tested $\tilde{\alpha}$ in equation~(\ref{eq:theta_bao}) as the normalized dispersion between the whole set correlation function and the one for the bin which contains $z_{eff}$. This bin is expected to be representative of the whole sample as described in section~\ref{sec:corre}. We can write it as
\begin{equation}\label{eq:alpha2}
    \tilde{\alpha}_{eff} (\theta)= \frac{\mathrm{w}_{eff}(\theta) - \mathrm{w}_{\delta z=0}(\theta)}{\sqrt{\int |\mathrm{w}_{eff}(\theta) - \mathrm{w}_{\delta z=0}(\theta)|^2 \mathrm{d} \theta}},
\end{equation}

\section{Results}\label{sec:results}

\subsection{Covariance matrix}\label{sec:covsec}

The correlation matrices of both mocks and data sets are shown in Fig.~(\ref{fig:covs}). 
The top left panel shows the correlation matrix of the BOSS1 sample, while the MultiDark Patchy Mocks results are shown on the top right one. Both plots show a higher correlation at smaller scales, from $\sim 1^{o}$ to $\sim 5^{o}$, and the data matrix is more noisy than the mocks, as one could expect. 

In the middle left panel, we can see that BOSS2 has more low correlation patches when compared to BOSS1, possibly due to the bigger number of objects and depth of this sample. Its mock counterpart shows much less noise, as expected, and similar over-correlated regions on both extremes of the angular separation. Additionally, the BOSS2 sample shows a high correlation between small and big scales that is not visible in BOSS1 and the mocks. 

Lastly, the results for eBOSS, the larger and deeper sample, are shown in the bottom panels of Fig~(\ref{fig:covs}). The data correlation matrix is clearly more noisy, and it does not show regions of stronger or weaker correlation, except for the diagonal. The EZ mocks appear to have a similar pattern to that of the other mocks, but there is a stronger correlation region on larger scales, unlike the other mocks.

As a means of validation, we computed the angular correlation function for each mock without our binning method. This full covariance equation is described below:
\begin{align}
    Cov^{full}_{ml} =\frac{n_{eff}^{-1}}{N-1} \displaystyle \sum\limits_{k=1}^{N=1000} [ \mathrm{W^{full}}_k(\theta_m) - \mathrm{\bar{W}^{full}}(\theta_m)] \nonumber \\ \times [\mathrm{W^{full}}_{k}(\theta_l)-\mathrm{\bar{W}^{full}}(\theta_l)].
    \label{eq:covmocks_full}
\end{align}

\begin{figure}
     \centering
     \includegraphics[width=.5\textwidth]{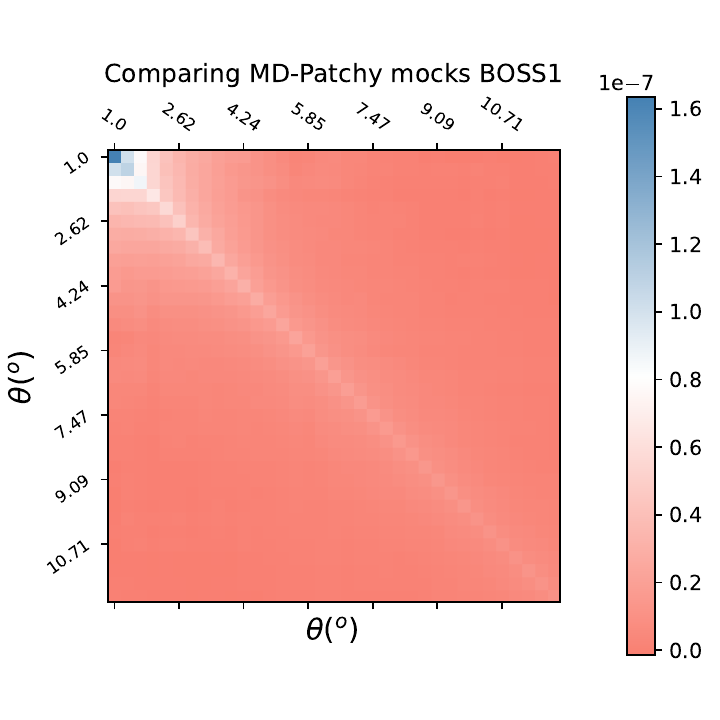}
     \caption{Comparison between our method and a covariance matrix without tomography.}
     \label{fig:boss1_compare}
 \end{figure}

\begin{figure}
     \centering
     \includegraphics[width=.5\textwidth]{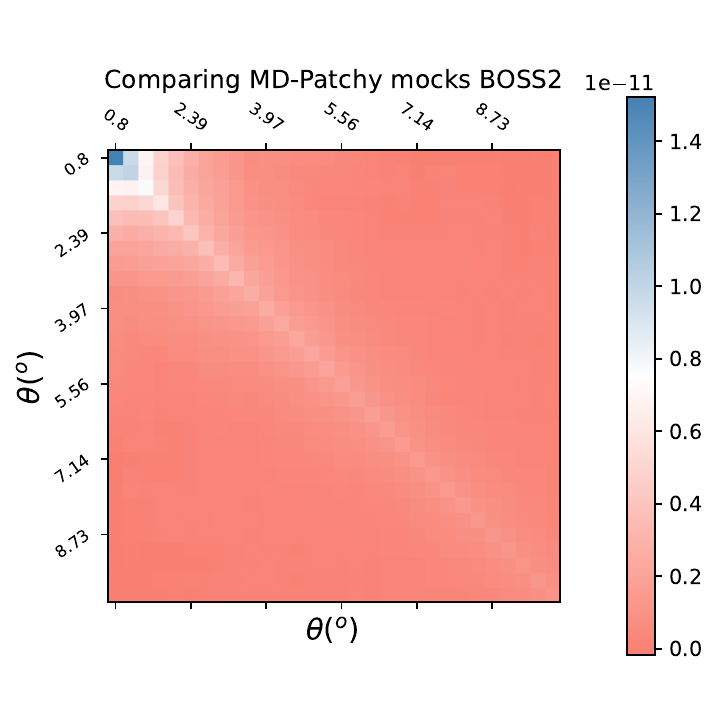}
     \caption{Comparison between our method and a covariance matrix without tomography.}
     \label{fig:boss2_compare}
 \end{figure}

\begin{figure}
     \centering
     \includegraphics[width=.5\textwidth]{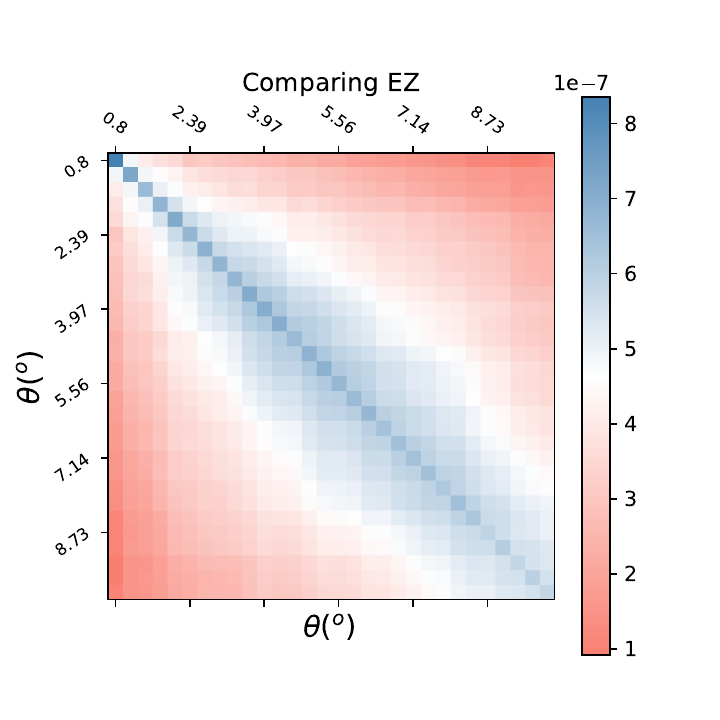}
     \caption{Comparison between our method and a covariance matrix without tomography.}
     \label{fig:eboss_compare}
 \end{figure}

\begin{figure}
\centering

\begin{subfigure}[b]{0.45\columnwidth}
    \includegraphics[width=\linewidth]{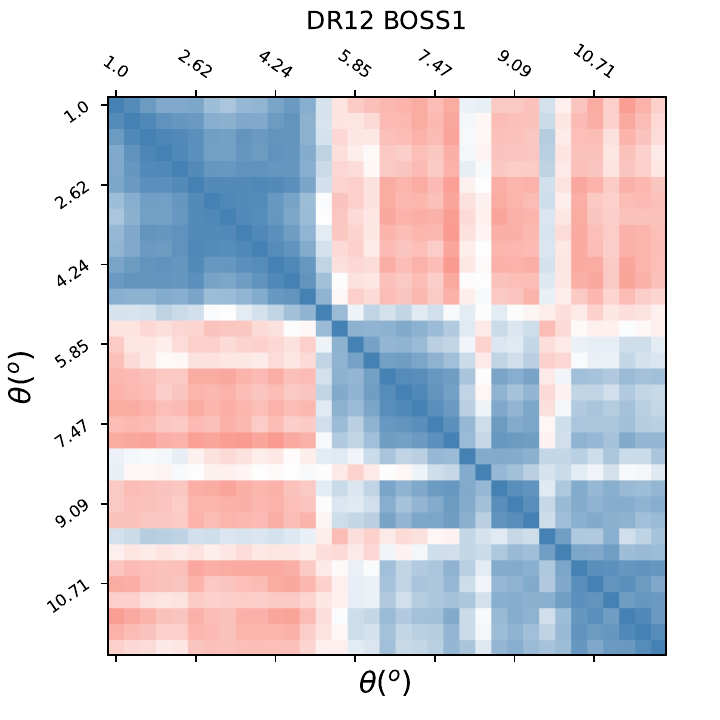}
\end{subfigure}
\begin{subfigure}[b]{0.45\columnwidth}
    \includegraphics[width=\linewidth]{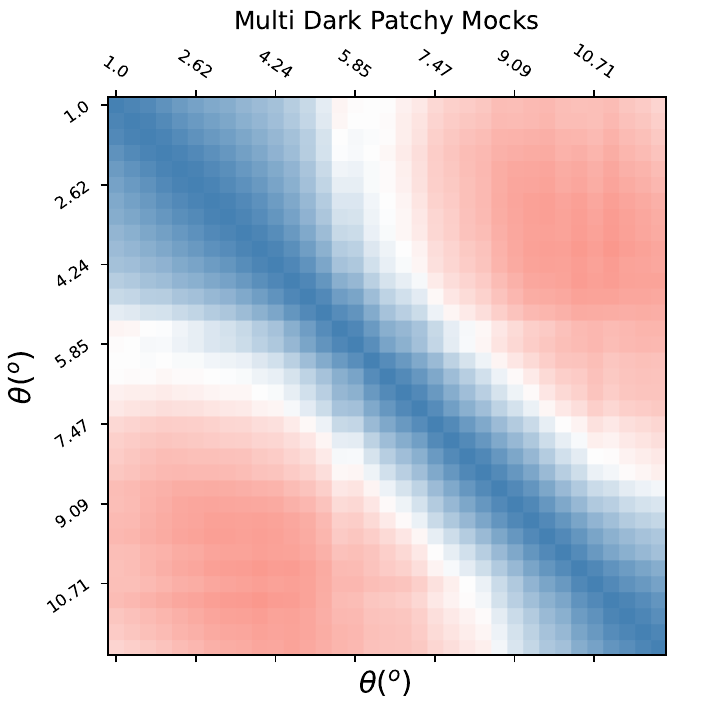}
\end{subfigure}

\begin{subfigure}[b]{0.45\columnwidth}
    \includegraphics[width=\linewidth]{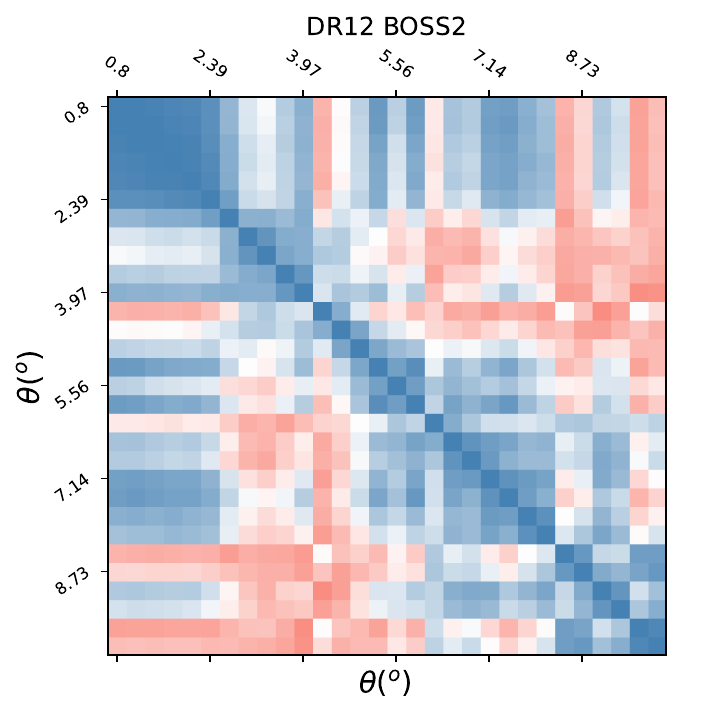}
\end{subfigure}
\begin{subfigure}[b]{0.45\columnwidth}
    \includegraphics[width=\linewidth]{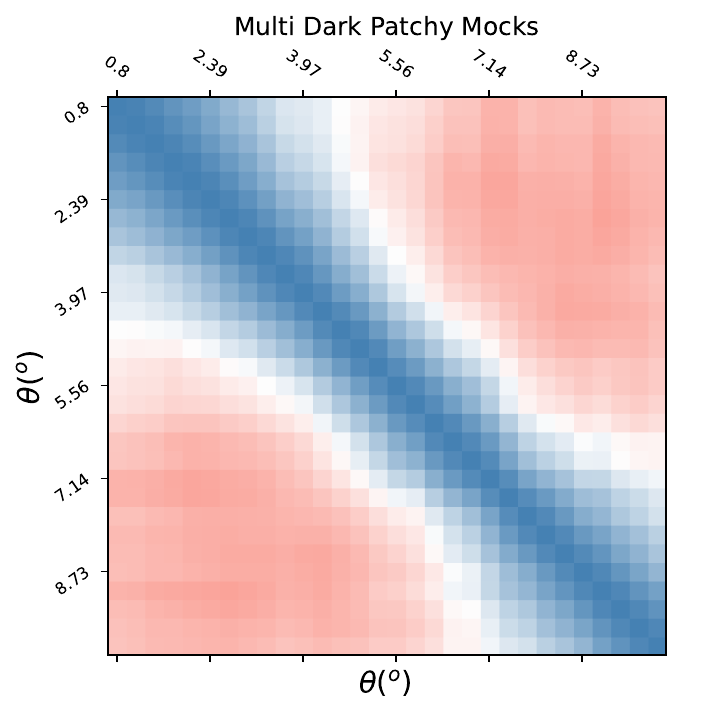}
\end{subfigure}
\begin{subfigure}[b]{0.45\columnwidth}
    \includegraphics[width=\linewidth]{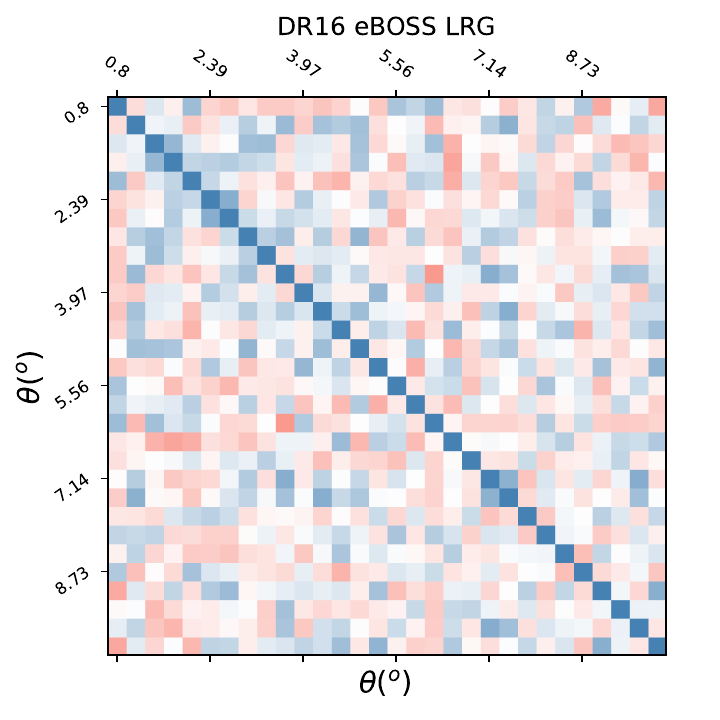}
\end{subfigure}
\begin{subfigure}[b]{0.45\columnwidth}
    \includegraphics[width=\linewidth]{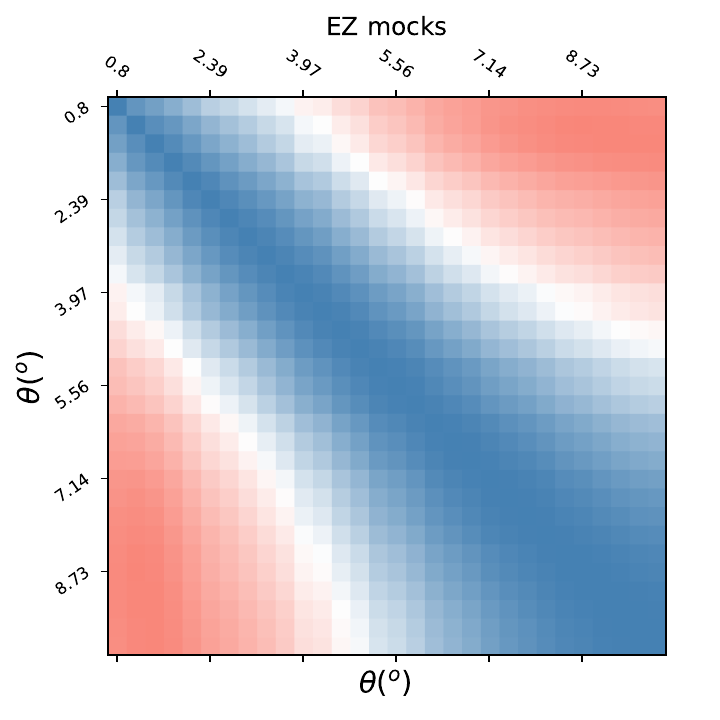}
\end{subfigure}
\begin{subfigure}{0.4\columnwidth}
    \includegraphics[width=\linewidth]{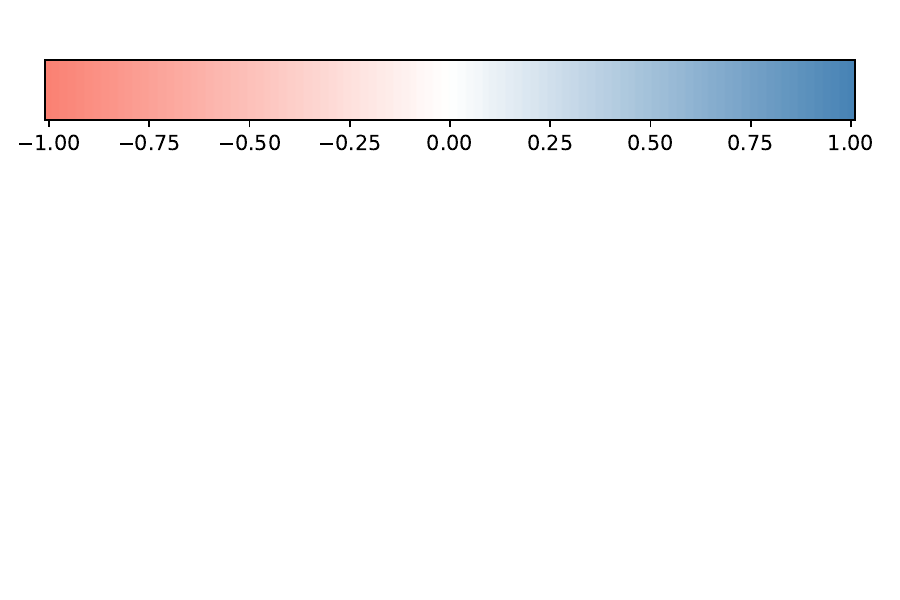}
\end{subfigure}
\caption{Correlation matrices relative to $\zeta_{ml}$, left column, and $C_{ml}$, right column. Negative correlation in salmon, a positive correlation in blue, and a zero correlation in white colour.}
\label{fig:covs}
\end{figure}

In figures \ref{fig:boss1_compare}, \ref{fig:boss2_compare}, and \ref{fig:eboss_compare}, we show $C_{ml} - Cov^{full}_{ml}$. The difference is of the order $10^{-7}$ for BOSS1 and eBOSS and $10^{-11}$ for BOSS2. For all samples, the difference is negligible near the BAO angular feature for the lower-z sample. However, this difference is larger for eBOSS, but still a small difference. This validates our method as a robust one.

We performed the minimum likelihood estimation with $Cov^{full}_{ml}$, the results are in table~(\ref{tab:params_full}), the results show that they agree in $1 \sigma$ with the constraints from $C_{ml}$.

\subsection{Polynomial fit}\label{sec:poly}
\begin{figure}
    \centering
    \resizebox{0.5\textwidth}{!}{\includegraphics{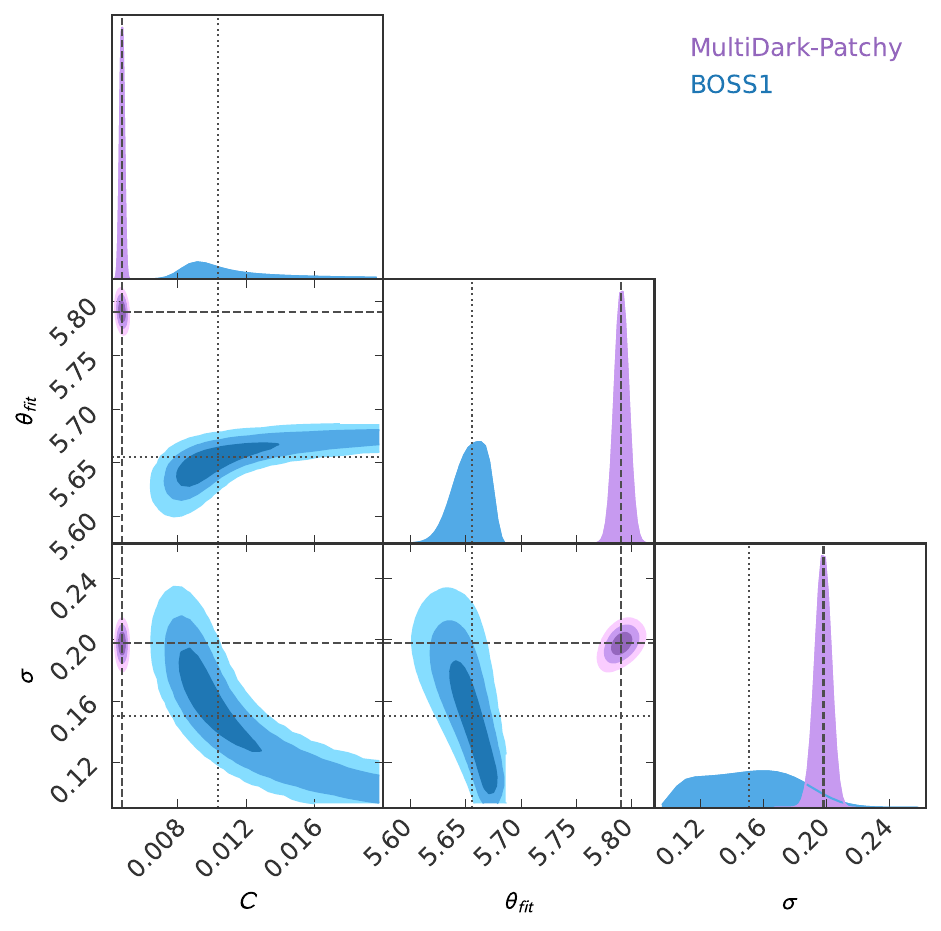}}
    \caption{Comparison between results for the BOSS1 using using $\zeta_{ml}$ in blue, and $C_{ml}$ in purple, using \texttt{pygtc} \citep{bocquet2016pygtc} for the parameters $C$, $\theta_{fit}$, and $\sigma$.}
    \label{fig:triangle1}
\end{figure}

\begin{figure}
    \centering
    \resizebox{0.5\textwidth}{!}{\includegraphics{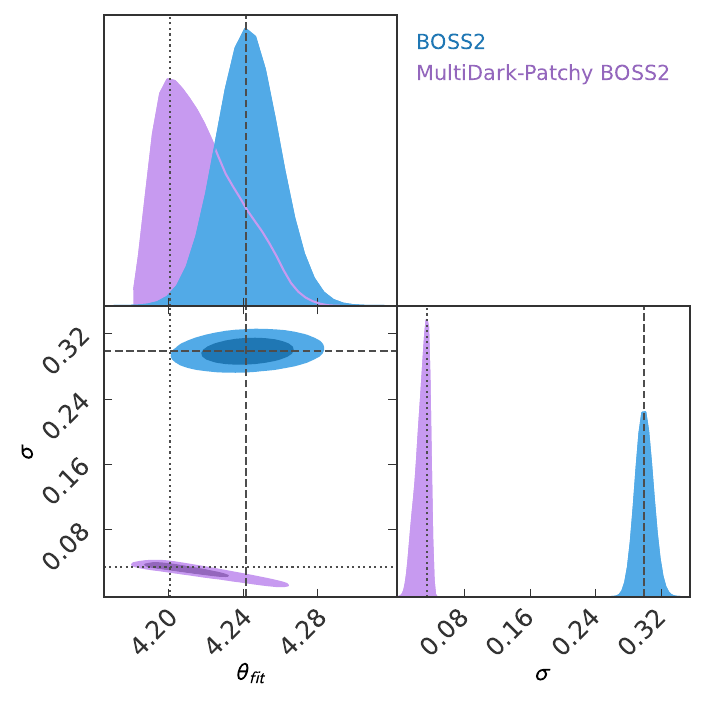}}
    \caption{Comparison between results for the BOSS2 using using $\zeta_{ml}$ in blue, and $C_{ml}$ in purple, using \texttt{pygtc} \citep{bocquet2016pygtc} for the parameters $C$, $\theta_{fit}$, and $\sigma$.}
    \label{fig:triangle2}
\end{figure}

\begin{figure}
    \centering
    \resizebox{0.5\textwidth}{!}{\includegraphics{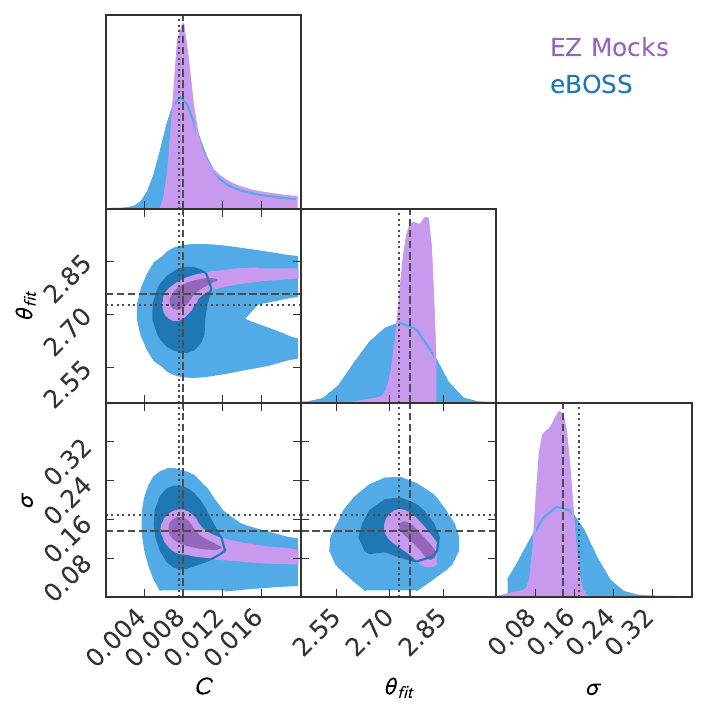}}
    \caption{Comparison between results for the eBOSS using using $\zeta_{ml}$ in blue, and $C_{ml}$ in purple, using \texttt{pygtc} \citep{bocquet2016pygtc} for the parameters $C$, $\theta_{fit}$, and $\sigma$.}
    \label{fig:triangle_eboss}
\end{figure}

After taking the steps described in the previous section, we get the estimated correlation function, where the results of the physical parameters of the fitting formula are compiled in Table~\ref{tab:params}. The comparison between mock and data fitting is shown in figures \ref{fig:triangle1}, \ref{fig:triangle2}, and \ref{fig:triangle_eboss}. The contour indicates the $68\%$, $95\%$, and $99.7\%$ confidence level regions, from darker to lighter colours.

\begin{table}
\centering
	\scriptsize	
	\caption{Best-fit parameters}
	\label{tab:params}
	\begin{tabular}{llll} 
            \hline\noalign{\smallskip}
		Sample & $C$ & $\theta_{fit}(^o)$ & $\sigma(^o)$\\
		\noalign{\smallskip}\hline\noalign{\smallskip}
		BOSS1($\zeta_{ml}) $&$0.010 \pm 0.001$& $5.66 \pm 0.02$ & $0.15 \pm 0.03$\\
            BOSS1($C_{ml}$) & $0.0047 \pm 0.0001$ & $5.79 \pm 0.006$ & $0.198 \pm 0.005$\\
            BOSS2($\zeta_{ml}$)& $0.002 \pm 0.0001$& $4.24 \pm 0.02$ & $0.30 \pm 0.01$ \\
            BOSS2($C_{ml}$) & $0.02 \pm 0.01$& $4.21 \pm 0.02$ & $0.030 \pm 0.009$\\
            eBOSS($\zeta_{ml}$)& $0.008 \pm 0.002$ & $2.72 \pm 0.09$ & $0.13 \pm 0.05$ \\
            eBOSS($C_{ml}$) & $0.009 \pm 0.001$ & $ 2.77 \pm 0.04$ & $0.12 \pm 0.03$\\
            \noalign{\smallskip}\hline
	\end{tabular}
\end{table}

We chose Gaussian priors for the physical parameters for all samples. But instead of using traditional random walkers, we used a mixture of moves in our code, in order to improve the performance of the MCMC walkers. 80\% of the moves are a "differential evolution" algorithm Differential Evolution MCMC (DEMCMC) by \cite{nelson2013run}, available in \texttt{emcee} as \texttt{DEMove}. The other 20\% of the moves are done with \texttt{KDEMove}.

Table \ref{tab:priors} shows the priors chosen for all cases. For all results shown in Figures \ref{fig:triangle1}, \ref{fig:triangle2} and \ref{fig:triangle_eboss}, and table \ref{tab:params}, we used the real data and changed the covariance as indicated. BOSS1 and BOSS2 results for the parameter $C$ did not agree at $\sim 5\sigma$ when we change the covariance from data to their respective mocks, while eBOSS does not show such issue.

eBOSS agrees within $1\sigma$ compared to their respective mocks for all parameters, as seen in Figure~\ref{fig:triangle_eboss}, with the mocks covariance presenting tighter constraints as expected due to the lower noise. The posteriors are not close to a Gaussian distribution, showing a long tail towards higher values of $C$.

BOSS1 did not show the same behaviour; the main reason is the size of the sample. It is reasonable to expect that the mocks do not include all possible nonlinear effects of the real LSS. This should be particularly important for lower redshift samples, where such effects will be greater. The method used here does not account for this depth dependence. These struggles with lower redshift and modelled solutions have been discussed extensively in the literature, especially in the 3D power spectrum case as seen in \cite{seo2008nonlinear} and \cite{xu20122}.

We see in Figure~\ref{fig:triangle1} that, despite the tension in the results, both $C_{ml}$ and $\zeta_{ml}$ show the BAO feature, with tighter 1$\sigma$ CL for $C_{ml}$ results. The first sample is affected by two peculiarities. As a consequence, we expect that the precision of the angular diameter distance will be affected.

BOSS2 presented better results based on $\zeta_{ml}$ (Figure~\ref{fig:triangle2}), while for $C_{ml}$  (Figure~\ref{fig:triangle2}) we obtained broader posteriors. This again reflects that this method does not account for nonlinear effects and that the mocks are not perfect idealizations of the LSS. 

\subsection{\texorpdfstring{$\theta_{BAO}$}{Lg} results}

The bias function from Eq.~(\ref{eq:normed_bias}) can be visualized as a scatter matrix between the redshift bins. Some bin distributions can be found in Appendix~\ref{ap:a}, in which we chose 10 bins close to the effective redshift to understand this bias relation. If the bias function is close to zero, the difference between the CFs of each bin and the whole set is supposed to be close to a symmetric distribution with mean zero.

\begin{figure}
\centering
    \includegraphics[width=0.45\textwidth]{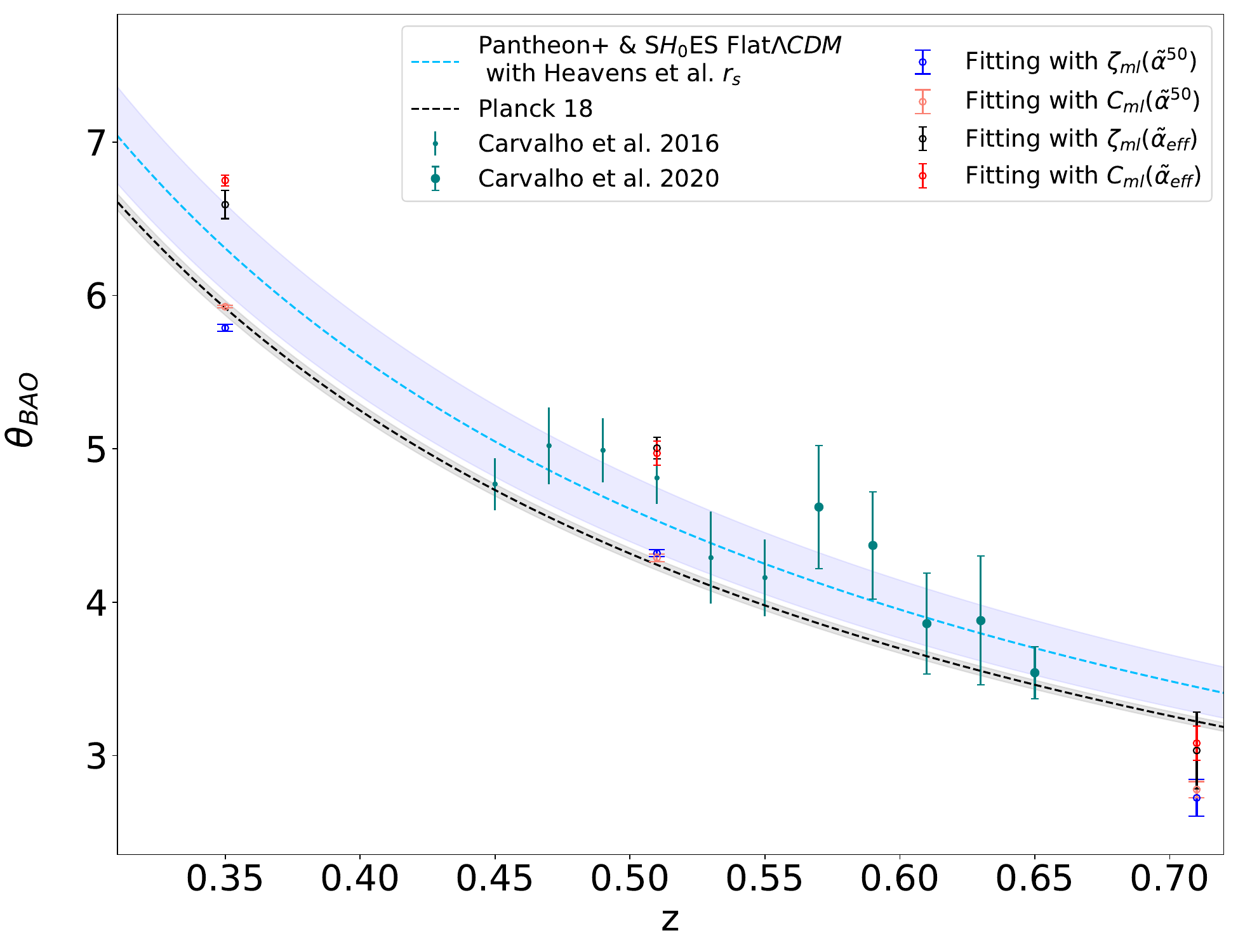}\\
    \includegraphics[width=0.45\textwidth]{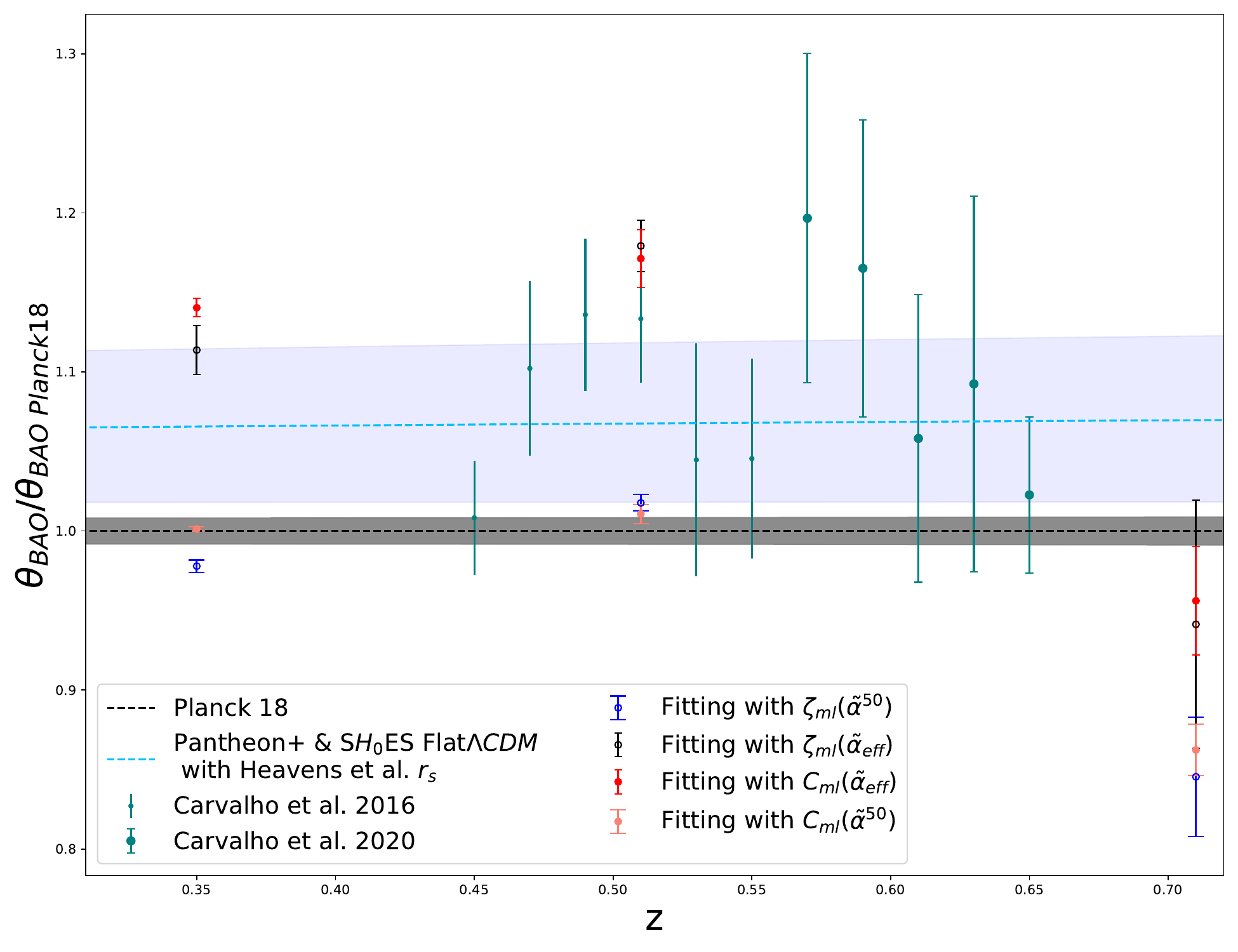}
    \caption{\textit{Top}$: \theta_{BAO}$ as a function of redshift with $\alpha$ from Eq.~(\ref{eq:alpha}) and (\ref{eq:alpha2}). Planck 18 \citep{aghanim2020planck} $\theta_{BAO}$ is shown as the black line and S$H_0$ES \citep{brout2022pantheon} as the dashed blue region. We kept the blue colour for the results with $\zeta_{ml}$ and salmon for $C_{ml}$. The shaded regions are the CL of Planck 18 in grey and S$H_0$ES in purple 68\% CL. \textit{Bottom}: Same result of the top panel divided by the Planck 18 one ($\theta_{BAO}/\theta_{BAO}^{Planck18}$).}
    \label{fig:theta_bao2}
\end{figure}

In Figure \ref{fig:theta_bao2}, top panel, we show the $\theta_{BAO}$ results from Eq.~(\ref{eq:theta_bao}). The sample covariance results are shown in blue, while the mock covariance results are in salmon. The black and red points represent the $\tilde{\alpha}_{eff}$ results of the sample covariance and mock covariance respectively. The sea-green points are results from \cite{carvalho2016baryon} and \cite{carvalho2020transverse}. 

As a means of comparison, we use two different cosmological parameter references to check our results. The first is the Planck 2018 Collaboration \citep{aghanim2020planck}, shown in black with a gray-shaded region. We considered the TT, TE, EE+lowE+lensing constraints. The second model is based on the Pantheon+ \& S$H_0$ES \citep{brout2022pantheon} Flat$\Lambda$CDM in light blue with a purple shaded region. The cosmological parameters can be found in Table~\ref{tab:models}. The shaded regions are the $68\%$ (darker) and $95\%$ (lighter) confidence levels (CL) for the results. The curves are plotted using the following relation as a function of redshift:
\begin{equation}
    \theta_{BAO}=\frac{r_s}{(1+z) D_A(z)},
\end{equation}
where $D_A(z)$ is the angular diameter distance as a function of redshift and $r_s=147.090 \pm 0.026$ Mpc is the sound horizon at drag epoch which is the Planck 18 \cite{aghanim2020planck} TT, TE, EE, lowE, lensing result for the black region. For S$H_0$ES fiducial model, we chose \cite{heavens2014standard} as the sound horizon, this gives $r_s=142.8 \pm 3.7$ Mpc. Figure \ref{fig:theta_bao2}, bottom panel, shows $\theta_{BAO}$ normalised by Planck 18 cosmology $\theta_{BAO}^{Planck18}$, in order to ease the comparison.

\begin{table}
\centering
    \scriptsize	
    \begin{tabular}{ccc}
        \hline\noalign{\smallskip}
         Parameters &Planck 18 & Pantheon+ \& S$H_0$ES \\
        \noalign{\smallskip}\hline\noalign{\smallskip}
         $\Omega_m$& $0.3153\pm0.0073$&$ 0.334 \pm0.018 $\\
         $\Omega_\Lambda$&$0.6847\pm0.0073$& $ 0.666 \pm 0.018$\\
         $H_0$(km s$^{-1}$ Mpc$^{-1}$)&$67.36\pm0.54$ &$73.6 \pm 1.1$\\
         \noalign{\smallskip}\hline
    \end{tabular}
    \caption{Parameters for Flat$\Lambda$CDM cosmology for Planck 18 \citep{aghanim2020planck} and Pantheon+ \& S$H_0$ES \citep{brout2022pantheon}.}
    \label{tab:models}
\end{table}

We see in Figure \ref{fig:theta_bao2}, top panel, that the highest $z_{eff}$ sample, eBOSS, has the larger error bars, this is due to the incompleteness of the sample. Since Eq.~(\ref{eq:CF}) weights the correlation function of each bin according to its random catalog size, this affects $\theta_{fit}$ and consequently $\theta_{BAO}$, as the number of galaxies per bin decreases with redshift for eBOSS, so the $n_{eff}$ is smaller compared to the other samples which increase the error bars. For all samples, the mock results have smaller error bars due to a larger number of galaxies per bin. Since we chose the same bin size ($\delta z = 0.002$) for all samples this results in fewer galaxies per bin for BOSS1 when compared to BOSS2 which leads to greater error bars for BOSS1. A summary of $\theta_{BAO}$ results is in \ref{tab:theta_bao_res}

\begin{table}
\centering
    \scriptsize	
    \begin{tabular}{cc}
        \hline\noalign{\smallskip}
         &$\theta_{BAO}(^o)$   \\
         \noalign{\smallskip}\hline\noalign{\smallskip}

          BOSS1 with $\tilde{\alpha}^{50}$ & $5.79\pm 0.02$ \\
          BOSS2 with $\tilde{\alpha}^{50}$ & $4.32\pm 0.02$ \\
          eBOSS with $\tilde{\alpha}^{50}$ & $2.72\pm 0.12$ \\
          BOSS1 mocks with $\tilde{\alpha}^{50}$ & $5.93 \pm 0.01$\\
          BOSS2 mocks with $\tilde{\alpha}^{50}$ & $4.29 \pm 0.03$\\
          eBOSS mocks with $\tilde{\alpha}^{50}$ & $2.79 \pm 0.05$\\
          BOSS1 with $\tilde{\alpha}_{eff}$ & $6.59 \pm 0.09$ \\
          BOSS2 with $\tilde{\alpha}_{eff}$ & $5.00 \pm 0.07$ \\
          eBOSS with $\tilde{\alpha}_{eff}$ & $3.03 \pm 0.25$ \\
          BOSS1 mocks with $\tilde{\alpha}_{eff}$ & $6.75 \pm 0.03$\\
          BOSS2 mocks with $\tilde{\alpha}_{eff}$ & $4.97 \pm 0.08$\\
          eBOSS mocks with $\tilde{\alpha}_{eff}$ & $3.08 \pm 0.11$\\
          \noalign{\smallskip}\hline
    \end{tabular}
    \caption{$\theta_{BAO}$}
    \label{tab:theta_bao_res}
\end{table}

It is shown in \ref{fig:theta_bao2}, top panel, that when using $\tilde{\alpha}_{50}$, BOSS1 and BOSS2 agree with Planck 18 in 2 $\sigma$ and 1 $\sigma$ respectively, while eBOSS is in tension with Planck 18. Using mocks covariance, the results for BOSS1 and BOSS2 both agree with Planck 18 results, eBOSS constraints with $C_{ml}$ also show a tension with Planck 18. This shows a tendency of the mocks' results to "reach" Planck 18 results.

When using $\tilde{\alpha}_{eff}$ the results change significantly. First, the error increases as a result of statistical loss, now we compare a single bin with $\mathrm{w}_{\delta=0}$. The $\theta_{BAO}$ now have higher values, BOSS1 and BOSS2 (in black) agree with S$H_0$ES results in 1$\sigma$ and $2\sigma$, respectively. eBOSS match Planck 18 in 1$\sigma$. BOSS2 has the same $z_{eff}$ as one of the bins from \cite{carvalho2016baryon} that, now, they agree in 1$\sigma$. We must stress that \cite{carvalho2016baryon,carvalho2020transverse} show larger error bars as a choice to avoid overlapping between bins.

The higher difference for the lower redshift samples can be explained by non-linearity influence, especially when we see the BOSS1 mocks' results (in red). It is hard to model all the necessary details of local galaxies clustering, as a result, $\theta_{BAO}$ is not in agreement with the sample covariance constraints.

\begin{table}
    \centering
    \scriptsize	
    \begin{tabular}{ccc}
        \hline\noalign{\smallskip}
         & $\theta_{BAO}^{20}-\theta_{BAO}^{50}$ & $\theta_{BAO}^{20}/\theta_{BAO}^{50}$\\
         \noalign{\smallskip}\hline\noalign{\smallskip}

          BOSS1 & $-6.97\times 10^{-4}$ & $99.98\%$\\
          BOSS2 & $-2.81\times 10^{-4}$ & $99.99\%$\\
          eBOSS & $-7.22\times 10^{-3}$ &$99.73\%$\\
          BOSS1 mocks & $-1.77 \times 10^{-3}$&$99.97\%$\\
          BOSS2 mocks & $-6.98 \times 10^{-3}$ &$99.98\%$\\
          eBOSS mocks & $-7.22\times 10^{-3}$&$99.73\%$\\
          \noalign{\smallskip}\hline

    \end{tabular}
    \caption{The difference between the 20$^{th}$ percentile and the 50$^{th}$ percentile.}
    \label{tab:theta_bao}
\end{table}

\cite{menote2022baryon} used the same data set, but chose a different approach from \cite{marra2019first} and a different selection cut than ours. Their method is based on \cite{marra2019first} as well as the present study but maintained the mocks as the only covariance matrix estimator. We had a similar pattern of $\theta_{BAO}$ as a function of the redshift. Like our results, many of their bins did not agree with Planck 18 cosmology. Moreover, their findings were capable of extending to cosmological parameter inference because we are just interested in the fiducial model-independent analysis itself. \footnote{In a recent draft of the Dark Energy Survey's final release, \cite{abbott2024dark}, their results show a larger discrepancy from Planck 18 cosmology compared to their Y3 results (figure 6). Moreover, in figure 10, there is a considerable shift in the BAO feature compared to Planck 18 results. }

It has been discussed by \cite{nunes2020cosmological} that 15 transversal BAO measurements combined with Planck 18 alleviates the tension between Planck 18 and \cite{riess2019large} Cepheid calibration.
\cite{nunes2020bao} also showed that combining the same measurements with Big Bang Nucleosynthesis (BBN) information \citep{adelberger2011solar} and strongly lensed quasars from H0LiCOW \citep{wong2020h0licow}, and BBN and cosmic chronometers (CC) \citep{moresco20166} agree in $1\sigma$ with $SH_0ES$ \citep{riess2019large} like our results using $\tilde{\alpha}_{eff}$. Still concerning tensions with Planck 18 cosmology, in \cite{bernui2023exploring}, the authors showed that the combination of Planck 18 and the 15 transversal measurements solves the $H_0$ tension by pushing to higher values of the Hubble parameter when considering an Interacting Dark Energy
(IDE) model.  Moreover, a considerable mismatch between Planck 18 cosmology and this study, \cite{menote2022baryon}, \cite{carvalho2016baryon}, and \cite{carvalho2020transverse} could be explained by the influence of non-linearity is much higher in the lower redshift samples as compared to CMB experiments. 

\section{Conclusions}\label{sec:conclusions}
We used BOSS and eBOSS LRG samples to obtain the angular feature of the BAO using thin bins to construct its covariance matrix. Each sample BOSS1, BOSS2, and eBOSS had 50, 100, and 200 thin redshift bins with $\delta z = 0.002$ width. We adapted \cite{sanchez2011tracing} and \cite{marra2019first} methodology and wrote an angular correlation function estimator through those thin bins using a weighting scheme based only on the random catalog.
We also performed the full analysis using both covariance matrices from mocks and real data.
We considered the random binning as the reference to the weight of each bin according to the number of galaxies Poisson distributed in the sky.

The comparison between mocks ($C_{ml}$) and sample covariance ($\zeta_{ml}$) best-fit showed that they agree at least for higher redshift. BOSS1 and BOSS2 showed disagreement with the physical parameters $C$ and $\sigma$. We must remind ourselves that the mocks are idealisations of the LSS and its non-linearity, but they may not represent the exact structure of the data, which could explain the disagreements in BOSS1 and BOSS2 results. We compared the method we proposed to traditional mocks without binning the data, the maximum difference is of the order $10^{-7}$ and is even smaller for the scale close to the feature of interest.

Furthermore, our approach is purely statistical, the way we divide the bins and the amount of galaxies for each bin also changes the covariance matrix. Therefore, one must divide the bins to hold statistical significance (many galaxies in one bin) and good correlation significance (as many bins as possible to a robust mean to represent the whole sample).

Considering statistical analysis, eBOSS performed better in our analysis, first, because it is a deeper sample. Second, it has more thin bins than the other samples, a closer approximation to a redshift-space correlation function, but losing statistical significance. The results showed agreement between the data and mock covariance estimation for all physical parameters.

We provided a model-free approach to estimate $\theta_{BAO}$ from $\theta_{fit}$ revisiting \cite{carvalho2016baryon}, we compute the correlation function of the whole sample and then find a bias compared to each bin correlation function. This is used to write a bias function $\tilde{\alpha}$ that shifts $\theta_{fit}$ closer to $\theta_{BAO}$. Instead of averaging over the bins, we get $\tilde{\alpha}(\theta)$ through the 20$^{th}$ and 50$^{th}$ percentile. 

The results were compared to the fiducial models of Planck 18 \citep{aghanim2020planck} and Pantheon+ \& S$H_0$ES \citep{brout2022pantheon} Flat$\Lambda$CDM with \cite{heavens2014standard} $r_s$. Regarding the percentile approach. our findings indicated the samples with lower redshift agree at least in $2 \sigma$ with Planck 18 results. eBOSS, on the other hand, did not agree with any of the fiducial cosmology described here.

When assuming $\tilde{\alpha}_{eff}$, the results change, we see that eBOSS matched Planck 18 results. BOSS1 and BOSS2 agree at least in 2$\sigma$ with Pantheon+ \& S$H_0$ES fiducial cosmology. We also found agreement with the same sample as BOSS2 from \cite{carvalho2016baryon} z=0.51 bin. This indicates that a loss of statistical significance interferes with how the measurement behaves provided that \cite{carvalho2016baryon} chose to lose many bins to avoid overlapping between them.

The mocks showed a tendency of getting $\theta_{BAO}$ closer to Planck 18 results than the real data. The only exception was BOSS1, the lower redshift sample which is more susceptible to non-linearity effects, something known in the literature as a challenging characteristic for obtaining survey mocks.

Next-generation surveys with large samples are suitable for the method. DESI, for instance, promises 8 million LRGs \citep{zhou2023target} with $0.4<z<1.0$, ideal to reduce the nonlinearity noise in our covariance matrix. These future samples should allow us to use thinner bins without reducing the number of galaxies per bin, which seems to be a key feature of this method. A further study is required to construct a test for cosmological models from that methodology, this would require a larger sample with higher redshift distribution, something interesting to future photometric surveys. This, however, also comes with the price of adding the photometric redshift uncertainties and even probability distribution functions of each object. 

\section*{Acknowledgements}

This work made use of the CHE cluster, managed and funded by COSMO/CBPF/MCTI, with financial support from FINEP and FAPERJ grant E-26/210.130/2023, and operating at the Javier Magnin Computing Center/CBPF.

The authors would like to thank the anonymous referee who provided useful and detailed comments.

Funding for the Sloan Digital Sky Survey IV has been provided by the Alfred P. Sloan Foundation, the U.S. Department of Energy Office of Science, and the Participating Institutions. SDSS acknowledges support and resources from the Center for High-Performance Computing at the University of Utah. The SDSS website is www.sdss.org.

SDSS is managed by the Astrophysical Research Consortium for the Participating Institutions of the SDSS Collaboration including the Brazilian Participation Group, the Carnegie Institution for Science, Carnegie Mellon University, Center for Astrophysics | Harvard \& Smithsonian (CfA), the Chilean Participation Group, the French Participation Group, Instituto de Astrofísica de Canarias, The Johns Hopkins University, Kavli Institute for the Physics and Mathematics of the Universe (IPMU) / University of Tokyo, the Korean Participation Group, Lawrence Berkeley National Laboratory, Leibniz Institut für Astrophysik Potsdam (AIP), Max-Planck-Institut für Astronomie (MPIA Heidelberg), Max-Planck-Institut für Astrophysik (MPA Garching), Max-Planck-Institut für Extraterrestrische Physik (MPE), National Astronomical Observatories of China, New Mexico State University, New York University, University of Notre Dame, Observatório Nacional / MCTI, The Ohio State University, Pennsylvania State University, Shanghai Astronomical Observatory, United Kingdom Participation Group, Universidad Nacional Autónoma de México, University of Arizona, University of Colorado Boulder, University of Oxford, University of Portsmouth, University of Utah, University of Virginia, University of Washington, University of Wisconsin, Vanderbilt University, and Yale University.

The massive production of all MultiDark-Patchy mocks for the BOSS Final Data Release has been performed at the BSC Marenostrum supercomputer, the Hydra cluster at the Instituto de Física Teorica UAM/CSIC, and NERSC at the Lawrence Berkeley National Laboratory. We acknowledge support from the Spanish MICINNs Consolider-Ingenio 2010 Programme under grant MultiDark CSD2009-00064, MINECO Centro de Excelencia Severo Ochoa Programme under grant SEV- 2012-0249, and grant AYA2014-60641-C2-1-P. The MultiDark-Patchy mocks was an effort led from the IFT UAM-CSIC by F. Prada’s group (C.-H. Chuang, S. Rodriguez-Torres and C. Scoccola) in collaboration with C. Zhao (Tsinghua U.), F.-S. Kitaura (AIP), A. Klypin (NMSU), G. Yepes (UAM), and the BOSS galaxy clustering working group.

PSF thanks Brazilian funding agency CNPq for PhD scholarship GD 140580/2021-2. RRRR thanks CNPq for partial financial support (grant no. $309868/2021-1$).

\appendix

\section{Bias function scatter matrices}\label{ap:a}

\begin{figure}
    \centering
    \resizebox{0.5\textwidth}{!}{\includegraphics{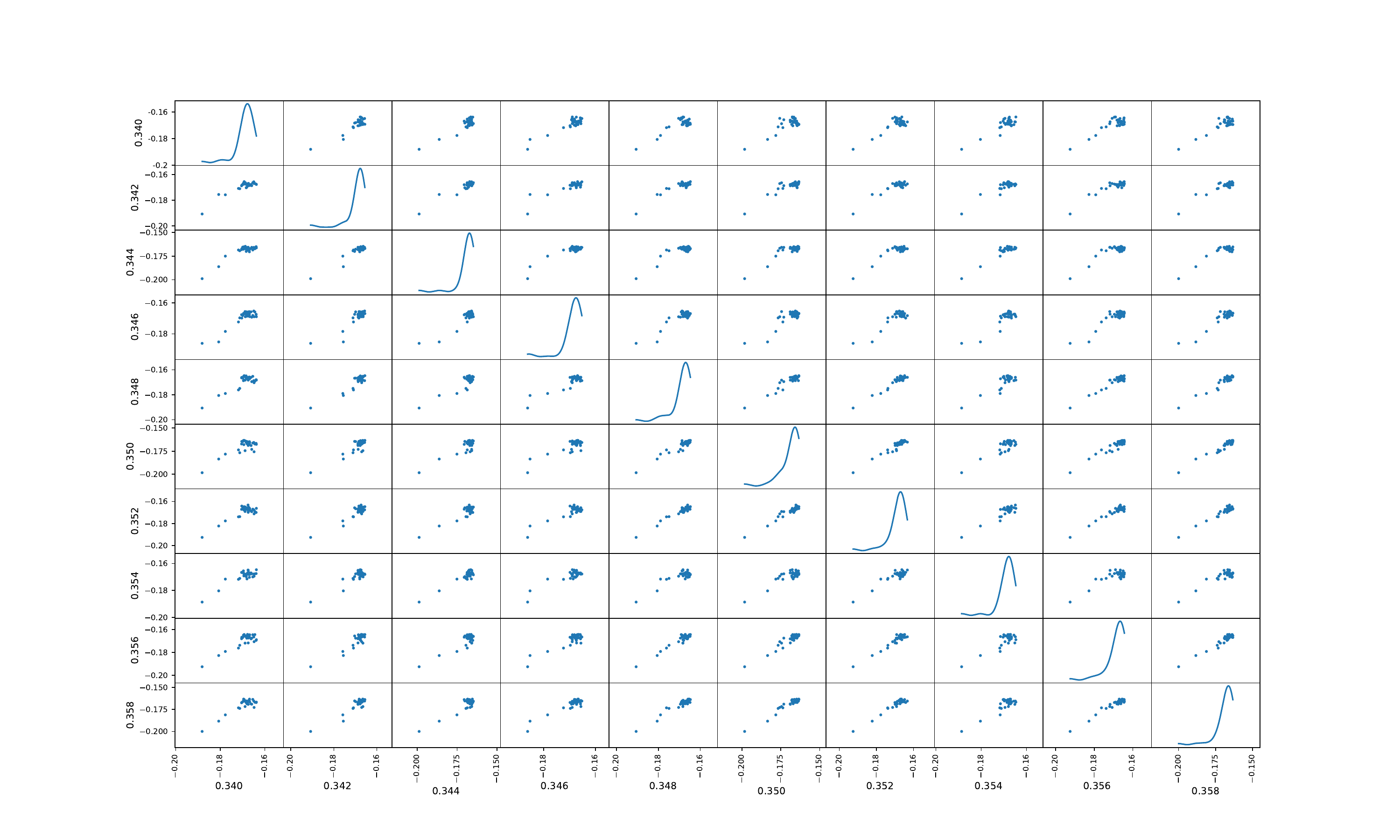}}
    \caption{BOSS1 scatter matrix of the bias relation between 10 neighbour bins. Both axes show the average $z$ for each bin.}
    \label{fig:disp1}
\end{figure}

\begin{figure}
    \centering
    \resizebox{0.5\textwidth}{!}{\includegraphics{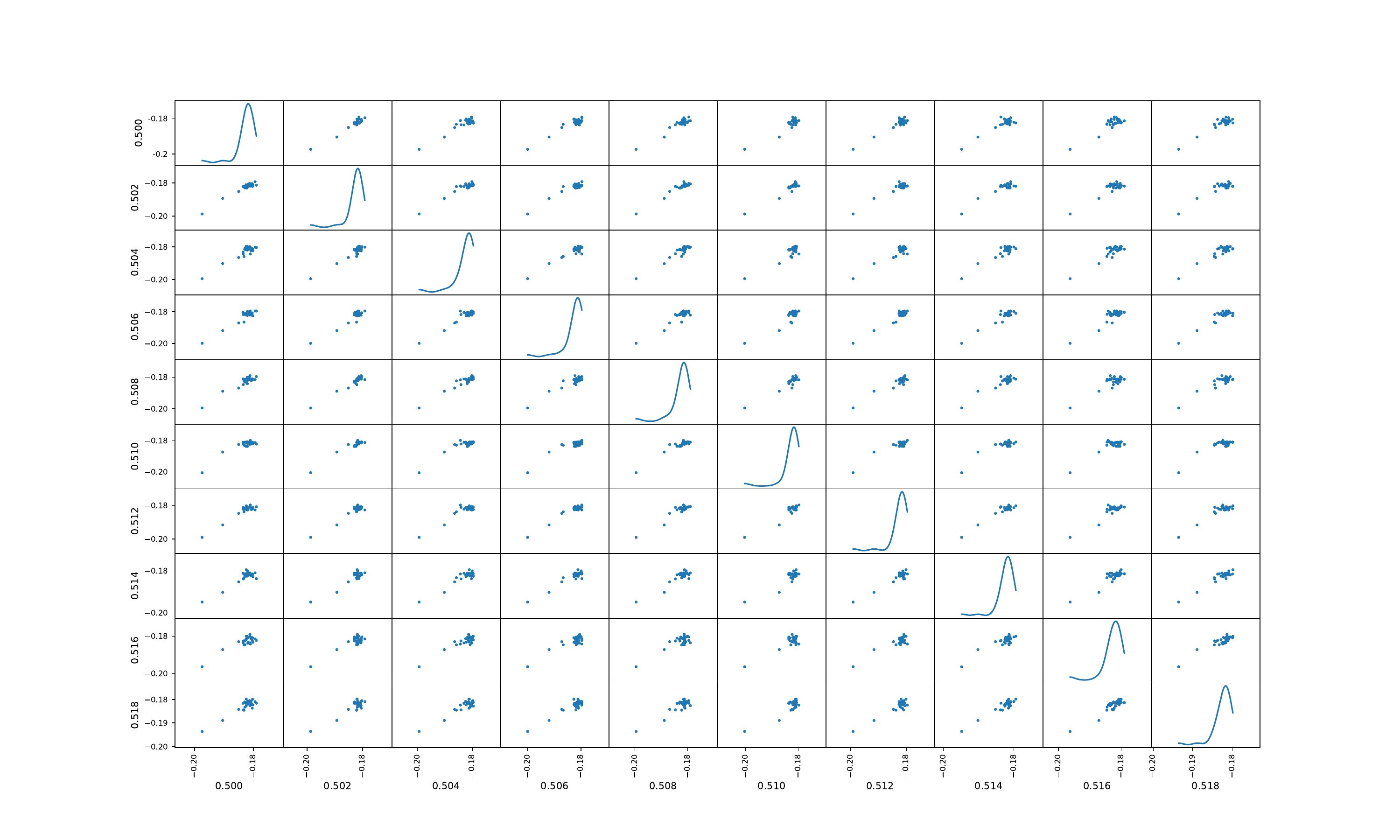}}
    \caption{BOSS2 scatter matrix of the bias relation between 10 neighbour bins. Both axes show the average $z$ for each bin.}
    \label{fig:disp2}
\end{figure}

\begin{figure}
    \centering
    \resizebox{0.5\textwidth}{!}{\includegraphics{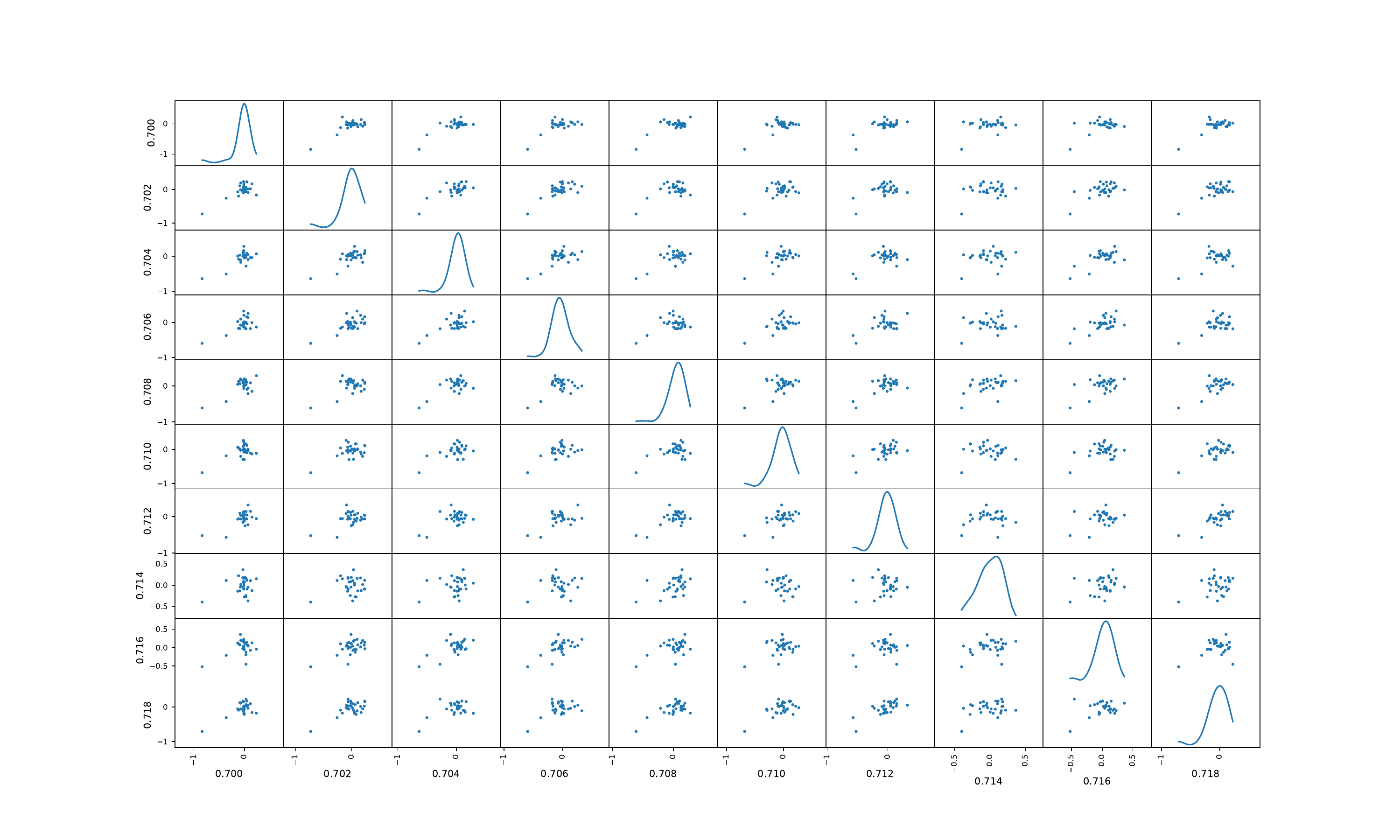}}
    \caption{eBOSS scatter matrix of the bias relation between 10 neighbour bins. Both axes show the average $z$ for each bin.}
    \label{fig:disp3}
\end{figure}

In figures \ref{fig:disp1}, \ref{fig:disp2}, and \ref{fig:disp3}, we present the scatter matrix of the bias function from Eq.~(\ref{eq:normed_bias})  for 10 bins close to the $z_{eff}$ for the three samples we chose to analyse. It is clear that BOSS1 and BOSS2 (figures \ref{fig:disp1}, \ref{fig:disp2}) show a high bias between the bins and whole set correlation function, in other words, they are not a symmetric distribution with mean zero.  eBOSS (\ref{fig:disp3}), on the other hand, maintains symmetric distributions even between bins distant from $z_{eff}$. This reflects the strong correlation between bins for low-z samples due to nonlinearities.

\begin{figure}
\centering
\begin{subfigure}[b]{.42\textwidth}\label{fig:boss2_zeta}
\centering
    \includegraphics[width=\textwidth]{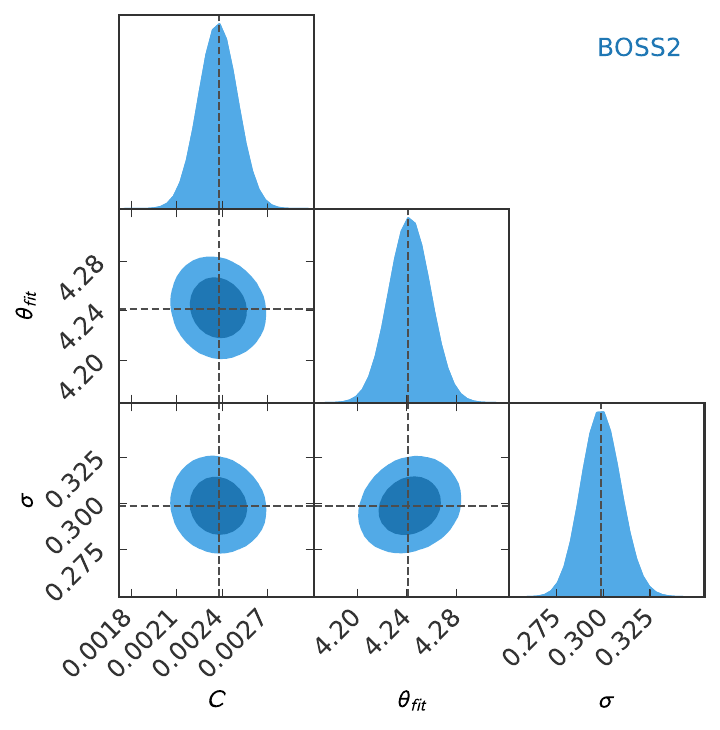}
    \caption{BOSS2 results with $\zeta_{ml}$.}
    \label{fig:triangle_0406}
\end{subfigure}
\begin{subfigure}[b]{.42\textwidth}\label{fig:boss2_c}
\centering
    \includegraphics[width=\textwidth]{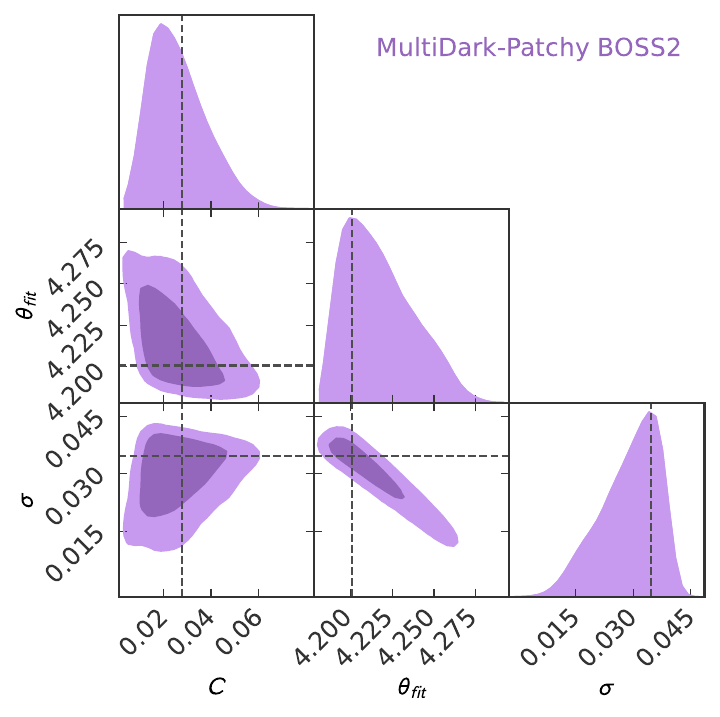}
    \caption{BOSS2 results with $\zeta_{ml}$.}
    \label{fig:triangle_0406_mocks}
\end{subfigure}
\caption{BOSS2 results using \texttt{pygtc} \citep{bocquet2016pygtc}.}
\end{figure}

\section{BOSS2 triangle plots}\label{ap:b}
In figures \ref{fig:triangle_0406} and \ref{fig:triangle_0406_mocks}, there are constraints for the separated physical parameters. The $C_{ml}$ results did not perform as well as the real data inference which agrees with the discussion of section~\ref{sec:poly}. The mocks should be a good representation, but this is more challenging for a shallower sample. The very different results are a combination of nonlinear effects representation for the mocks and our method not being successful in fixing such issues.

\section{Comparing with full mocks}
 In table \ref{tab:params_full}, we show the constraints with the full mocks samples.

\begin{table}
\centering
	\caption{Best-fit parameters using $Cov^{full}_{ml}$.}
	\label{tab:params_full}
	\begin{tabular}{llll} 
            \hline
		Sample & $C$ & $\theta_{fit}(^o)$ & $\sigma(^o)$\\
		\hline

		BOSS1 &$  0.0065\pm 6$& $5.798 \pm 2.0 $ & $2.7857 \pm 2 $\\
                   & $\times10^{-10}$ &$\times10^{-8}$ &$\times 10^{-8}$ \\
            BOSS2 & $(9.11 \pm 0.00005) $& $4.2\pm 15$ & $1.83\times 10^{-1} \pm 1 $ \\
                & $\times 10^{-4}$&$\times 10^{-7}$ & $\times 10^{-7}$\\
            eBOSS & $(8.090 \pm 0.0001) $ & $2.74 \pm 6 $ & $( 162.301\pm 0.004) $ \\
                & $\times 10^{-3}$& $\times 10^{-5}$&$\times 10^{-3}$ \\
            \hline
	\end{tabular}
\end{table}

\newpage
\bibliographystyle{mnras}
\bibliography{references}

\end{document}